\documentclass[aip,jap,reprint]{revtex4-1}
\usepackage{graphicx}
\usepackage{ulem}
\usepackage{amssymb}
\usepackage{amsmath}
\usepackage{wasysym}
\begin{document}
%\draft                    

%\flushbottom
%\twocolumn[
%\hsize\textwidth\columnwidth\hsize\csname @twocolumnfalse\endcsname

\newcommand{\be}{\begin{equation}}
\newcommand{\ee}{\end{equation}}
\newcommand{\bea}{\begin{eqnarray}}
\newcommand{\eea}{\end{eqnarray}}
\newcommand{\Tbar}{{\bar{T}}}
\newcommand{\En}{{\cal E}}
\newcommand{\K}{{\cal K}}
\newcommand{\GC}{{\cal \tt G}}
\newcommand{\Lop}{{\cal L}}
\newcommand{\DB}[1]{\marginpar{\footnotesize DB: #1}}
\newcommand{\q}{\vec{q}}
\newcommand{\kt}{\tilde{k}}
\newcommand{\Lopn}{\tilde{\Lop}}
\newcommand{\noi}{\noindent}
\newcommand{\ovn}{\bar{n}}
\newcommand{\ovx}{\bar{x}}
\newcommand{\ovE}{\bar{E}}
\newcommand{\ovV}{\bar{V}}
\newcommand{\ovU}{\bar{U}}
\newcommand{\ovJ}{\bar{J}}
\newcommand{\calE}{{\cal E}}
\newcommand{\ovphi}{\bar{\phi}}
\newcommand{\zt}{\tilde{z}}
\newcommand{\rt}{\tilde{\rho}}
\newcommand{\tth}{\tilde{\theta}}
\newcommand{\ttz}{\tilde{z}}
\newcommand{\ttr}{\tilde{\rho}}
\newcommand{\nuv}{{\rm v}}
\newcommand{\ck}{{\cal K}}
\newcommand{\cc}{{\cal C}}
\newcommand{\ca}{{\cal A}}
\newcommand{\cb}{{\cal B}}
\newcommand{\cg}{{\cal G}}
\newcommand{\ce}{{\cal E}}
\newcommand{\fn}{{\small {\rm  FN}}}

\title{Approximate universality in the electric field variation on a field-emitter tip in the presence of space charge}
\vskip 0.15 in

\author{Raghwendra Kumar}
\affiliation{
Bhabha Atomic Research Centre,
Mumbai 400 085, INDIA}

\author{Gaurav Singh}
\affiliation{
Bhabha Atomic Research Centre,
Mumbai 400 085, INDIA}
\affiliation{Homi Bhabha National Institute, Mumbai 400 094, INDIA}
\author{Debabrata Biswas}\email{dbiswas@barc.gov.in}
\affiliation{
Bhabha Atomic Research Centre,
Mumbai 400 085, INDIA}
\affiliation{Homi Bhabha National Institute, Mumbai 400 094, INDIA}

%\pacs{85.45.-w}{}
%\pacs{03.65.Sq}{}
%\pacs{03.65.Xp}{}
%\pacs{52.59.Sa}{}

\begin{abstract}
  The electric field at the surface of a curved emitter is necessary to calculate the
  field emission current. For smooth parabolic emitting tips where space charge is negligible,
  variation of the electric field at the surface is known to follow the generalized cosine law.
  Here we investigate the validity of the cosine law in the regime where space charge due to
  emitted electrons is important. Particle-in-Cell (PIC) simulations with an emission algorithm
  based on the cosine law is employed for this study. It is shown that if $E_P$ and $E_L$ be the
  field at the apex of tip with  and without space charge respectively, then for  $\vartheta=E_P/E_L \geq 0.9$, the average relative deviation of the electric field from the cosine law is less than $3\%$ over the endcap. Thus, an emission scheme based on cosine law may be used in PIC simulations of field emission of electrons from curved emitter tips in the weak space charge regime. The relation between $\vartheta$ and normalized current $\zeta$ for curved emitters in this regime is also investigated. A linear relation, $\vartheta=1 - \delta \zeta$ (where $\delta$ is a constant), similar to that obtained theoretically for flat emitting surfaces is observed but the value of $\delta$ indicates that the extension of the
theory for curved emitters may require incorporation of the field enhancement factor.        
\end{abstract}

\maketitle

\section{Introduction}
\label{sec:intro}
Field emission is one of the important mechanisms of electron emission from the surface of a material. The theory of field emission was first formulated by Fowler and Nordheim (FN) in the year 1928\cite{FN,Nordheim}. Since then several important aspects of field emission theory have been studied by researchers\cite{burgess,murphy,jensen2003,forbes2006,jensen_book,FD2007,DF2008,KX2015,jensen2019,db_parabolic,db_rr_2019,rr_db_2021,db_rr_2021} and it is now increasingly being accepted that the curvature-corrected Murphy-Good formalism currently offers the best description of field emission\cite{forbes2019a,gated}.

The expression for the emitted current density at a point on the surface of emitter based on Murphy-Good\cite{murphy} formulation reads as:

\begin{equation}
  J({\bf r}) = \frac{1}{t_F^2}\frac{A_{FN}}{\phi} (E_l({\bf r}))^2 \exp\left(-{B_{FN}}{v_F} {\phi}^{3/2}/E_l({\bf r})\right).
    \label{eq:MG}
\end{equation}

\noi
Here 
$A_\fn~\simeq~1.541434~{\rm \mu A~eV~V}^{-2}$,
$B_\fn~\simeq 6.830890~{\rm eV}^{-3/2}~{\rm V~nm}^{-1}$ are the conventional Fowler-Nordheim constants
while $v_F = 1 - f + (f/6) \ln f$ and $t_F = 1 + f/9 - (f/18)\ln f$ are corrections due to the image
charge potential with 
$f  \simeq  c_S^2~E_l({\bf r})/\phi^2$ where $c_S$ is the Schottky constant and
$c_S^2 = 1.439965~ \text{eV}^2 \text{V}^{-1} \text{nm}$. $E_l({\bf r})$ refers to the local electric field on the emitter surface, while $\phi$ is the work function of the emitter material which we shall consider to be 4.5eV.
Eq.~(\ref{eq:MG}) is best suited\cite{db_rr_2019,db_rr_2021} when the apex radius of curvature $R_a > 100$nm and we shall stick to this regime through the rest of the paper.

It is well known that field emission of electrons from a surface requires a high electric field, typically $>3$ V/nm for an emitter having work function $\phi = 4.5$eV. The local field enhancement at a curved surface due to geometrical effects makes it possible to generate such high local fields even at relatively lower applied macroscopic electric field. The local field enhancement factor $\gamma$ is the ratio of local electric field to the asymptotic electric field  $E_0$, away  from emitting surface\cite{edgcombe2002,forbes2003,db_fef,db_ss_fef}.  It may be noted that the maximum field enhancement occurs at the apex of the curved emitting structure\cite{db_fef}. 

To get the total current emitted from the tip, knowledge of the electric field at all the points on emitting surface is important. For axially symmetric emitters with locally parabolic tips, it has been shown that the magnitude of the local electric field $E_l$ at the surface of the emitter $z = z(\rho)$ follows a `universal' cosine law \cite{cosine1,cosine2}

\bea
E_l(\tth) & =  & {E_a} \cos{\tilde \theta} \label{eq:cos1} \\
\cos{\tilde \theta} & = & \frac{z/h}{\sqrt{(z/h)^2+(\rho/R_a)^2}}. \label{Eq:cosine2}
\eea

\noi
In the above, $E_a$ denotes the electric field at the apex, $h$ is height of the emitter, $R_a$ is
radius of curvature at apex of the emitter and $\rho$ is the distance of point on the surface from
the axis of emitter. Thus, the net current obtained by
integrating Eq.~(\ref{eq:MG}) over the emitter surface $z = z(\rho)$, can be expressed in terms of the apex electric
field $E_a$ and the apex radius of curvature $R_a$ since the electric field at any point on the surface
is related to apex  electric field using Eqns.~(\ref{eq:cos1}) and (\ref{Eq:cosine2}). Thus, if the cosine
law is valid, it is necessary to determine only the apex electric field accurately in order to predict
the net field emission current.

Note that the locally parabolic assumption $z \approx h - \rho^2/(2R_a)$ includes all generic emitters with a smooth endcap, although the extent upto which parabolicity holds depends on the nature of the endcap. Thus, for a hemispherical endcap for instance, the parabolic approximation holds close to the apex and can therefore be used only for low electric fields. A hemiellipsoid endcap  on the other hand, with height greater than $3R_a$ follows the parabolic approximation further away from the apex and can be used to predict the net current at higher local field strengths.  

The cosine law greatly simplifies the job of calculating the net emitted current. It can
also be profitably used for emission-modeling in Particle-In-Cell\cite{Birdsall} (PIC)
simulation\cite{ms_hybrid}. The standard approach in PIC codes is
to solve Maxwell's equations, obtain the field on the emitting surface and then
use Eq.~(\ref{eq:MG}) to get the emission current from each surface area element.
The charge per time-step can then be evaluated and computational
particles launched from each area element normal to the surface. In case
of a curved emitter tip however, the accurate evaluation of charge emitted at each time step from
sufficiently fine surface elements on the emitting object, adds to the
computational resources enormously in a
Finite Difference Time Domain (FDTD) based PIC simulation. 
An alternate emission method  in a PIC code\cite{ms_hybrid}, is to use the generalized
cosine law\cite{cosine1,cosine2} to arrive at the net emission current
using only the apex electric field and the apex radius of curvature.  
A knowledge of the electric field variation on the emitter endcap (the generalized cosine law)
also enables us to cast the distribution of electrons emitted
in a geometry-independent universal
form, in terms of the generalized angle $\tth$, the total energy and the normal energy\cite{tth}.
Thus, while the alternate approach still requires a solution to the Poisson equation,
only the apex field needs to be sufficiently accurate to capture the total emission current.
The charge per time step can then be divided into a desired number of computational particles and the particles
can be emitted using distributions for the angle $\tilde \theta$ and conditional distributions
for the normal energy $\mathcal{E}_N$ and total energy $\mathcal{E}_T$.

The cosine law can thus reduce the computational burden in PIC simulations. It
has been shown to hold for different kind of emitter geometries\cite{cosine2}. Although, it
was originally derived for the anode far away from the emitter apex, it holds for the
anode in close proximity to the emitter\cite{anodeprox} as well as in the presence of other emitters such as in an array or a random arrangement of emitters\cite{db_rr_2020}. In fact, numerical
simulations show that it is also valid for gated emitters\cite{gated} and multiscale
geometries \cite{db_schottky,ms_hybrid}.

To the best of our knowledge, it is not known whether the cosine law holds in the presence of space charge in the diode region. This is crucial for the determination of the net emitted current using the apex electric field and the subsequent use of the distribution based particle-injection discussed above. We have attempted to address this question in this communication. Due to the high sensitivity of field emission current on the local electric field at the emitting surface, a small change in electric field can result in a large change in the emitted current.  The presence of charges in the diode reduces the field on the emitter surface resulting in a reduction in emitted current\cite{barbour,jpv2006,Forbes08,chen,Jensen15,Tofason15}. This type of limitation in current due to space charge is different from space charge limited emission in case of thermal or explosive field emitters. This phenomenon, where relatively weak space charge limits the current by affecting field emission from the emitting surface, has been referred to as field-emitted vacuum space charge (FEVSC) by Forbes\cite{Forbes08} or  space charge affected field emission (SCAFE) by Jensen et. al\cite{Jensen15}. We shall investigate the validity of cosine law in this regime.

The paper is organized as follows. We first describe the method employed to test the validity of cosine law  in section \ref{sec:method}. The geometry  of the diode and other details about simulation are given in section \ref{sec:geom}. This is followed by numerical results and analysis in section \ref{sec:result}. Concluding remarks will be presented in section \ref{sec:conclusions}.

\section{Methodology}
\label{sec:method}

The question of validity of the cosine law in the presence of space charge is investigated in this communication using the Particle-in-Cell (PIC) simulation technique. PIC simulation involves (i) update of electric and/or magnetic field on computational grid by solving Maxwell equations (ii) movement of charged particles under the influence self generated and/or imposed electric and magnetic field and (iii) assignment of charge and current density due to charged particles at the computational grid to update the field. In general, PIC codes employ finite difference method. This may be attributed to robustness of algorithms for assignment of charge and current density in finite difference approach. 

\begin{figure}[hbt]
  \begin{center}
 \vskip -2.50cm
%\hskip  3.5cm
\hspace*{0.75cm}
\includegraphics[scale=0.5,angle=0]{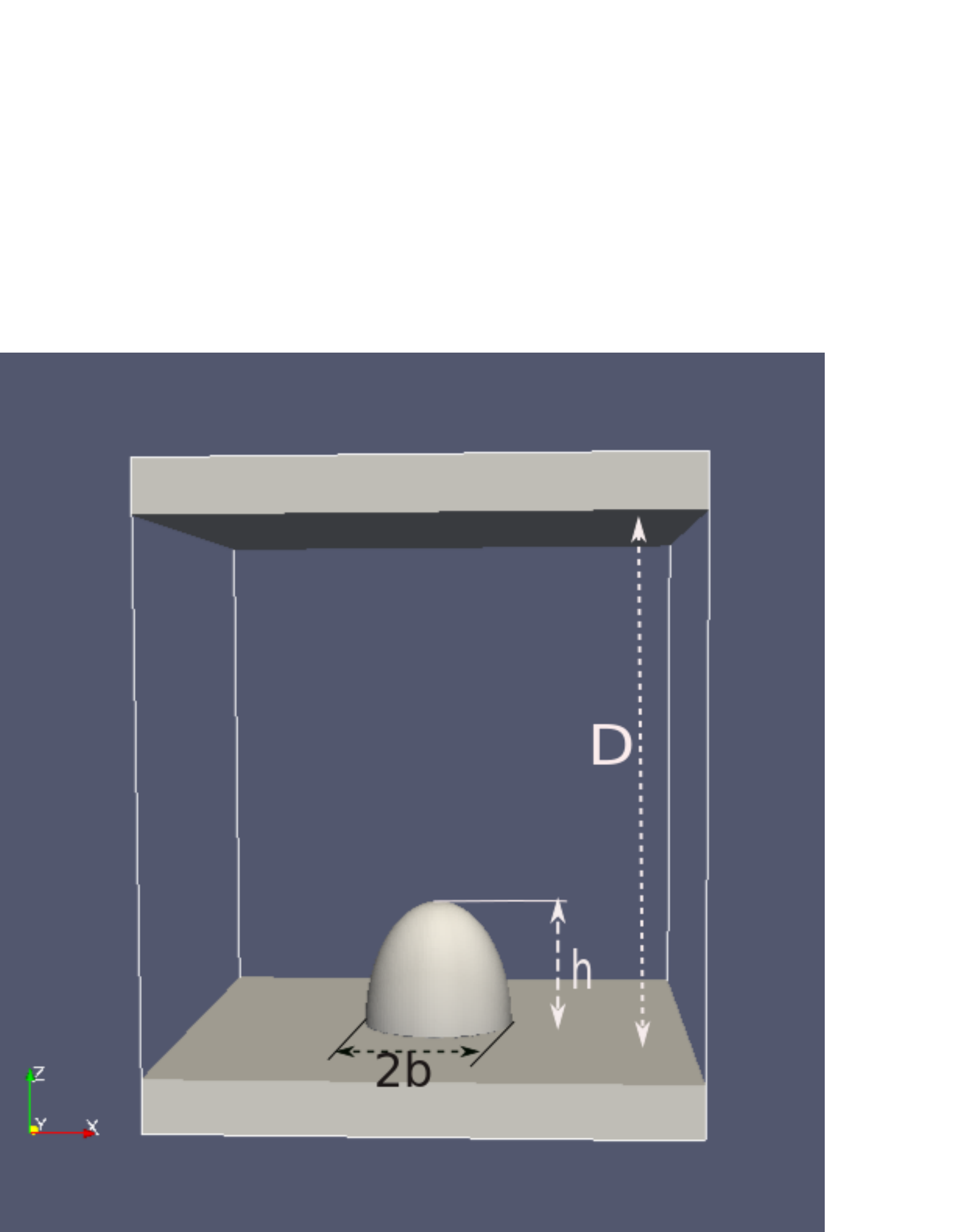}
\vskip -0.25cm
\caption{Geometry of the diode modeled in PASUPAT. $D$ is diode gap, $h$ and $b$ are the height and  base radius of the hemi-ellipsoidal emitter respectively.}
\label{fig:geom_diode}
\end{center}
\end{figure}

A regular computational grid poses challenge in handling surfaces of curved boundaries. As the computational cells at the curved boundaries are cut into two medium by a boundary surface, special algorithms are required to smoothen the field\cite{cut-cell-ES}. Emission of charged particle using Eq.~\ref{eq:MG} from such cells also requires special care\cite{loverich,edelen}. 

Furthermore, in order to model the electron flow in a diode properly, the computational time step should be small compared to the transit time of the particle. Note that the ballistic transit time is given by $T_{transit} = \sqrt{2mD/eE_0}$, where $D$ is the diode gap-length. For a micrometer size emitting tip, an accurate estimation of the field requires the grid to be sufficiently fine. Thus, the area of the element at the emitting surface  in a cut-cell is very small. As a result, the amount of charge emitted from such an element in a computational time step turns out to be less than the charge of an electron. This poses a problem as far as convergence with respect to the number of computational particles is concerned since the charge per particle moves further away from the charge of the physical particles being modeled.   

Recently, an emission algorithm based on cosine law of field variation at emitting tip was used in a study of electron beam transport across a gated diode \cite{ms_hybrid}. The space charge effect on the emission was negligible in that study and hence the cosine law was assumed to be valid. The emission algorithm involves: (i) estimation of charge emitted from the emitting tip\cite{db_parabolic} using a combination of Eq.~\ref{eq:MG} and the cosine law (ii) divide  the charge into a number of computational macro-particles (iii)  emit the particles from area lying between $\tilde \theta$ to ${\tilde \theta}+\Delta {\tilde \theta}$  by sampling $\tilde \theta$ using distribution function derived in Ref. [\onlinecite{db_parabolic}].  The distribution function reads:

\begin{equation}
f({\tilde \theta}) = 2 \pi R_a^2 {\frac{\sin{\tilde \theta}}{\cos^2 {\tilde \theta}}} J({\tilde \theta}) {\frac{{\tilde z}^{3}}{C_1}} C_0
\label{Eq:dist}
\end{equation}

%f({\tilde \theta},\mathcal{E_N},\mathcal{E_T}) 
\noindent
where the quantities ${\tilde z}$, $C_0$ and $C_1$ are defined in the appendix \ref{sec:appen}. For sharp emitters ($h/R_a >>1$), further simplification can be made and the distribution function assumes the form:

\begin{equation}
f({\tilde \theta}) \simeq 2 \pi R_a^2 {\frac{\sin{\tilde \theta}}{\cos^2 {\tilde \theta}}} J({\tilde \theta}).
\end{equation}

As mentioned in Ref.~[\onlinecite{ms_hybrid}], first ${\tilde \theta}$ is sampled from the distribution described above. The value of the normal energy $\mathcal{E_N}$ and total energy $\mathcal{E_T}$ are thereafter obtained using the conditional distributions $f(\mathcal{E_N}|{\tilde \theta})$ and $f(\mathcal{E_T}|\mathcal{E_N},{\tilde \theta})$ respectively. These distributions can be arrived at using the joint distribution  $f({\tilde \theta},\mathcal{E_N},\mathcal{E_T})$ \cite{db_parabolic}. Once the angle $\tth$, normal energy and total energy are calculated, velocity components in local co-ordinate system are obtained using suitable transformations.

Our rationale in investigating the validity of cosine law in the presence of space charge is as follows. As there is no charge in the diode gap at the start of the simulation, the cosine law is valid and hence charged particles (computational particle in PIC) are emitted using the distribution function described above. As space charge builds up in the diode region, the apex electric field changes. The field solver of the PIC code calculates the net field at the apex of the emitting tip, and this field in turn is used for further emission using the distribution function. The electric field at the surface of emitter is recorded at different times and its variation along the surface is studied to see whether it follows the cosine law. If the deviation from the cosine law is large, it can be concluded that the emission algorithm is invalid. On the other hand, if the deviation is small even after achieving the steady state, the cosine law holds and the emission algorithm is valid for the parameters of the diode.

\section{Geometry and Simulation Details}
\label{sec:geom}
We have carried out simulation study using the PIC code PASUPAT \cite{PoP_PASUPAT,ms_hybrid}. A parallel plate diode with hemi-ellipsoidal emitter on the cathode plate has been considered in the simulation as shown in Fig~\ref{fig:geom_diode}. Different values of diode gap-length $D$, emitter height $h$ and base radius $b$ have been considered (see table \ref{tab:table-1}). The hemi-ellipsoidal emitter is placed at the center of the cathode plate. The orientation of the emitter is along the $z$ direction. The anode plate is kept at $z=D$ from the cathode plate. Periodic boundary conditions are imposed along $x$ and $y$ boundaries which are kept sufficiently away from the hemi-ellipsoidal emitter to simulate an isolated emitter.

The number of computational cells along $x$, $y$ and $z$ were typically taken to be $\text{nx}=128$, $\text{ny}=128$ and $\text{nz}=256$.  We have varied the grid, time-step and particle number to ensure convergence. The Cut-Cell module\cite{cut-cell-ES} of PASUPAT has been deployed to model curved emitter. Since an electrostatic simulation has been carried out, a Multigrid Poisson solver\cite{multigrid} has been used to update the electric field. The emission module described in Sec.~\ref{sec:method} is used to estimate the net field emission current, the number of computational particles to be emitted in a time step, the distribution on the endcap with respect to the angle $\tth$ and finally the
initial velocities of the particles. 

\section{Results} 
\label{sec:result}

In this study three geometries have been considered. Parameters $(D,h,b)$  of the respective geometries 1,2, and 3 are specified in  Table \ref{tab:table-1}.

\begin{table}[]
  \begin{center}
    \caption{Geometric parameters considered in the study}
    \vskip 0.1 in
    \label{tab:table-1}
    \begin{tabular}{|c|c|c|c|}
      \hline
     ~~Geometry~~  & ~~$D$($\mu$m)~~   & ~~$h$($\mu$m)~~ & ~~$b$($\mu$m)~~\\
      \hline
      1  &1.0      &0.2515      &0.15\\
       \hline
      2  &10.0     &2.515       &1.5 \\
       \hline
      3  &100.0    &25.15       &15 \\
      \hline
    \end{tabular}
\end{center}
\end{table}

\begin{figure}[hbt]
  \begin{center}
% \vskip -2.50cm
%\hskip  3.5cm
\hspace*{-0.7cm}
\includegraphics[scale=0.35,angle=0]{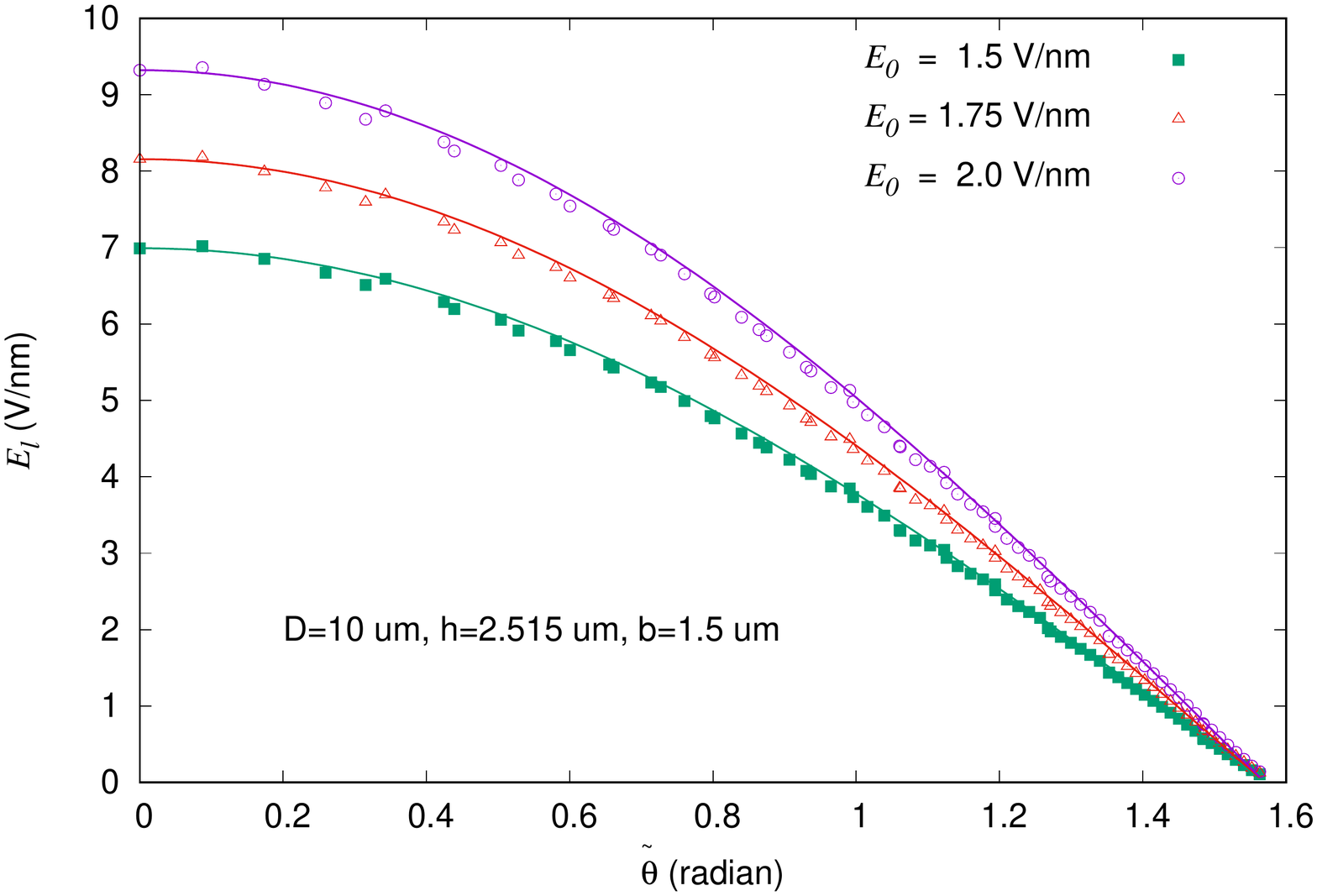}
\vskip -0.25cm
\caption{Variation of electric field on the surface with generalized angle ${\tilde \theta}$ for three different values of asymptotic electric field $E_0$ at the start of simulation ($t = 0$) for Geometry 2. Points are values obtained using PASUPAT PIC simulation while the solid curves correspond to ${E_a} \cos{\tilde \theta}$. Clearly, the electric field follows the cosine law in the absence of space charge.}
\label{fig:cosine_law1}
\end{center}
\end{figure}

We shall first consider Geometry 2 and compare the magnitude of electric field on the surface of the emitter at the start of simulation ($\text{nt}=0$, $\text{nt}$ being the simulation time step) in Fig. \ref{fig:cosine_law1}. The points (circle, square and triangle) are obtained by the PIC code PASUPAT while the curves represent the local field $E_l$ evaluated using the cosine law $E_l(\tth) = E_a \cos\tth$ with $E_a$ computed using PASUPAT. Solid squares, solid triangles and open circles represent surface fields for $E_0= 1.5$ V/nm, $E_0= 1.75$ V/nm and $E_0= 2.0$ V/nm respectively. Initially, there is no particle inside the diode gap and hence there is no space charge effect. It is well established that in the absence of space charge, the electric field at the surface of hemi-ellipsoid follows cosine variation\cite{cosine1,cosine2}. Fig.~\ref{fig:cosine_law1} shows that fields obtained using PASUPAT agree reasonably well with the cosine law. A similar study for Geometry 1 and 3 also shows that the electric field on  the emitter surface follows the cosine law  at the start of simulation.

\begin{figure}[hbt]
  \begin{center}
% \vskip -2.50cm
%\hskip  3.5cm
\hspace*{-0.7cm}
\includegraphics[scale=0.35,angle=0]{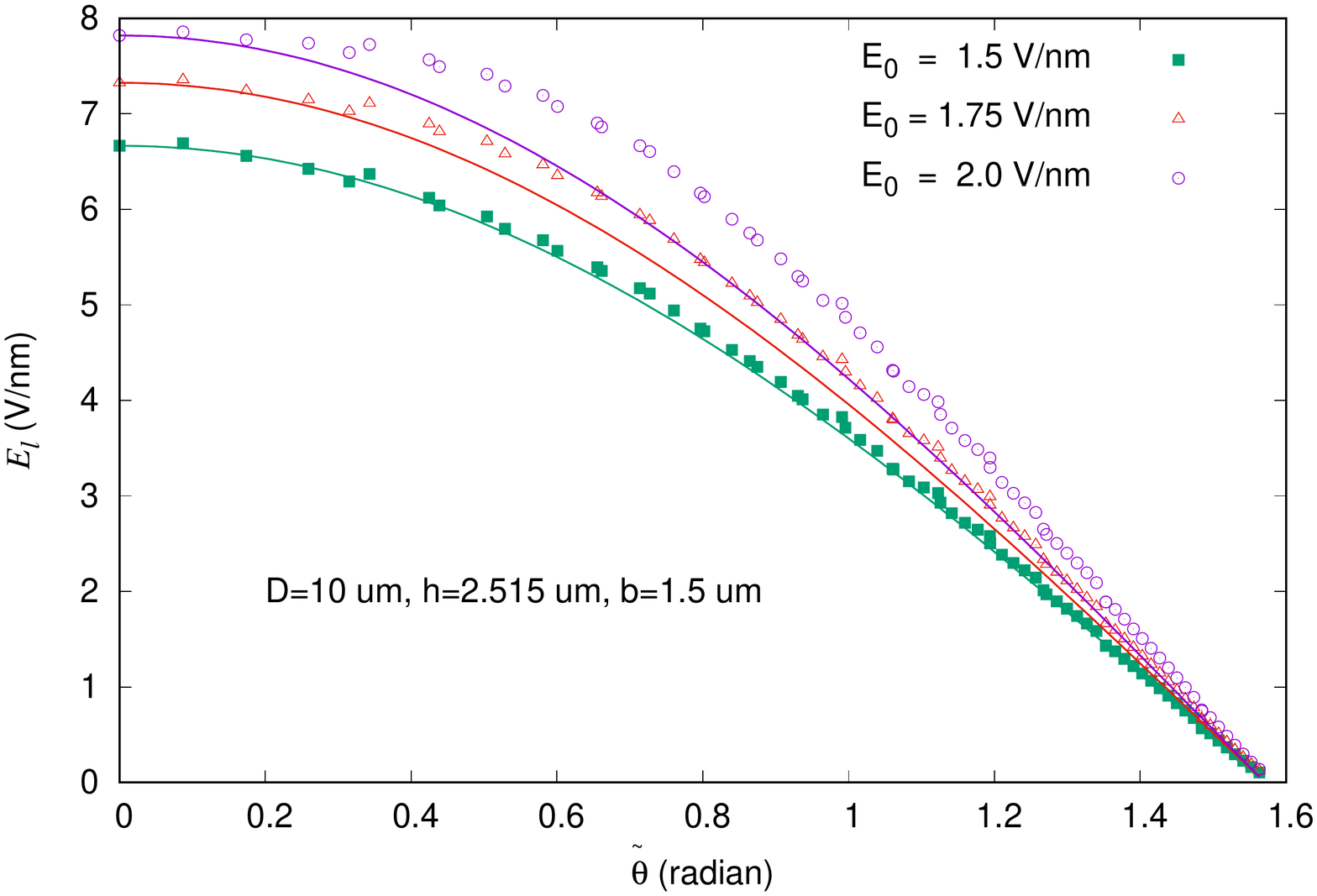}
\vskip -0.25cm
\caption{Variation of electric field on the surface with generalized angle ${\tilde \theta}$ averaged approximately over one transit time using fields obtained every 500 time steps. Three different values of asymptotic electric field $E_0$ are considered. Points are values obtained using PASUPAT PIC simulation while solid lines correspond to ${E_a} \cos{\tilde \theta}$. For lower ${E_0}$, the variation of local field follows the cosine law closely, while as $E_0$ increases, deviation from cosine law is significant. }
\label{fig:cosine_law2}
\end{center}
\end{figure}

\begin{figure}[hbt]
  \begin{center}
 \vskip -2.0cm
%\hskip  3.5cm
\hspace*{-0.7cm}
\includegraphics[scale=0.35,angle=0]{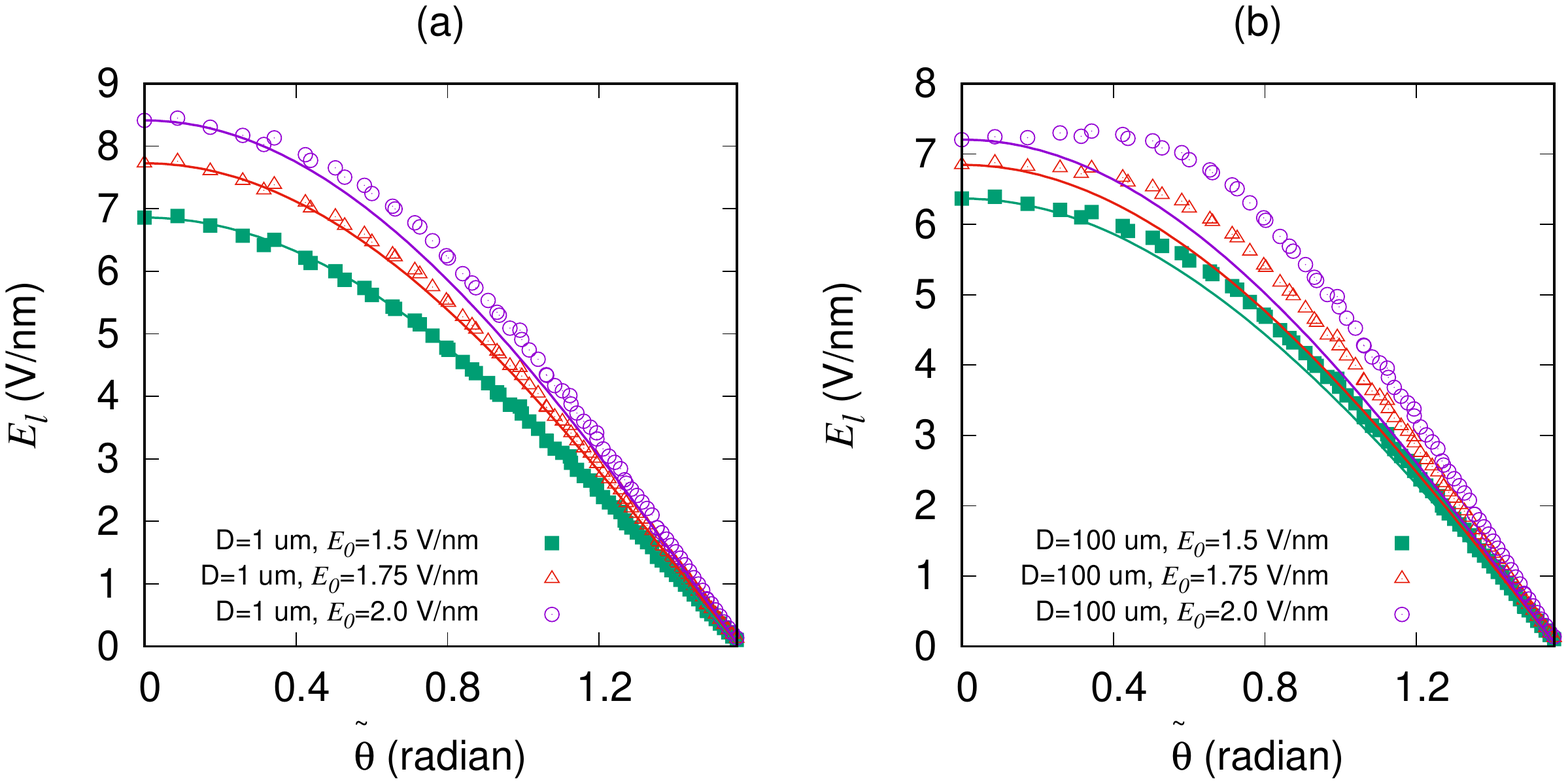}
\vskip -1.5cm
\caption{Variation of time-averaged electric field on the surface with generalized angle ${\tilde \theta}$ for three different values of asymptotic electric field $E_0$  (a) $D=1 \mu$m, $h=$0.2515 $\mu$m and $b=$0.15 $\mu$m (Geometry 1) and (b)  $D=100 \mu$m, $h=$25.15 $\mu$m and $b=$15 $\mu$m (Geometry 3).  Points are values obtained using PASUPAT PIC simulation and solid lines correspond to ${E_a} \cos{\tilde \theta}$. For lower $E_0$, the field variation follows the cosine law in both cases, while as $E_0$ increases, deviation from the cosine law is significant for Geometry 3.}
\label{fig:cosine_law3}
\end{center}
\end{figure}

We have calculated the average percentage relative error in the PIC field with respect to local field obtained using the cosine law as follows:

\be
\Xi = {\frac{1}{N_{\text{p}}}} \sum_{\tilde \theta_{i}=0}^{\tilde \theta_{R_a}}{\frac{|~E_a \cos {\tilde \theta_{i}} - E_{\text{PIC}}(\tilde \theta_i)~|}{E_a \cos {\tilde \theta_i}}\times 100},
\ee   

\noindent
where both $E_a$ and $E_{\text{PIC}}$ are calculated using PASUPAT. $E_a$ is the field at the apex of the tip ($\tilde \theta = 0$) while $E_{\text{PIC}}(\tilde \theta_i)$ denotes the field on a point on the surface specified by $\tilde \theta_i$. $N_{\text{p}}$ is the number of points between $\rho=0$ to $\rho_m = 3R_a/4$ and $\tilde \theta_{R_a}$ corresponds to generalized angle corresponding to $\rho_m$. For all the cases, $\Xi \approx 1.2$ \% at the start of the simulation\cite{surface_field}.

As the simulation proceeds in time, space charge builds up in the diode region and the field at the emitting surface reduces. After about one ballistic transit time, the average emission current settles down to a lower value compared to the start of simulation. We have studied the variation of the field at the emitter surface, averaged over the fields obtained at intervals of 500 time steps between the second and third transit time.  The number of simulation time steps in a single transit time is approximately 3000. Fig.~\ref{fig:cosine_law2} shows the variation of the magnitude of the electric field on the surface of emitter for three asymptotic fields $E_0$, 1.5 V/nm, 1.75 V/nm and 2.0 V/nm in case of Geometry 2. As before, points are obtained using the PIC code PASUPAT while the curves represent the cosine law. Solid squares, solid triangles and open circles represent surface fields for $E_0= 1.5$ V/nm, $E_0= 1.75$ V/nm and $E_0= 2.0$ V/nm respectively. It may be seen that the field variation agrees very well with the  cosine law for $E_0$=1.5 V/nm while the deviations increase at the largest field. 

Apart from the applied field, the validity of the cosine law in the presence of space charge also depends on the size of diode region. This is evident from Fig.~\ref{fig:cosine_law3} which shows the field variation for Geometry 1 (Fig.~\ref{fig:cosine_law3}a) and Geometry 3 (Fig.~\ref{fig:cosine_law3}b) which differ in size by two orders of magnitude. As before, the variation is averaged using the fields obtained at intervals of 500 time steps between the second and third transit time. As in Fig.~\ref{fig:cosine_law2}, solid squares, solid triangles and open circles represent surface fields for $E_0= 1.5$ V/nm, $E_0= 1.75$ V/nm and $E_0= 2.0$ V/nm respectively for both (a) and (b). Solid curves represent the corresponding cosine variation.  It may be noted that the local fields are identical in all three cases at the start of the simulation. However Geometry 1 has the smallest transit time and emission area while Geometry 3 has the largest. As a result, Geometry 3 has a larger accumulation of space charge compared to Geometry 1, thus leading to a larger deviation from the cosine law for the same value of applied field. Geometry 1 on the other hand is compatible with the cosine law even at higher values of $E_0$.

A measure of space charge is defined as $\vartheta=E_P/E_L$, where $E_P$ is electric field in the presence of space charge and $E_L$ is that obtained by solving Laplace equation\cite{Forbes08}. Here we take $E_P$ as apex electric field averaged from the second to third transit time as before while $E_L$ is the apex field at $\text{nt}=0$. In order to quantify the effect of space charge on the validity of cosine law, the deviation $\Xi$ for different values of $\vartheta$ is shown in
table \ref{tab:table-2}.

\begin{table}[h]
  \begin{center}
    \caption{Average error for different cases in the presence of space charge}
    \vskip 0.1 in
    \label{tab:table-2}
    \begin{tabular}{|c|c|c|c|}
      \hline
     ~~Geometry~~  &~~$E_0~\text{(V/nm)}$~~       &~~$\vartheta$~~ &~~$\Xi$(\%)~~\\
      \hline
         &1.5      &0.981     &0.57\\
      1  &1.75     &0.947     &1.06 \\
         &2.0      &0.902     & 2.62  \\
      \hline
         &1.5      &0.953      & 0.96 \\
      2  &1.75     &0.898      &3.14 \\
        &2.0      &0.839      & 5.88 \\
      \hline
        &1.5      &0.910      & 2.81 \\
      3  &1.75     &0.839      & 6.32 \\
         &2.0      &0.773      & 10.07\\
      \hline
    \end{tabular}
  \end{center}
\end{table}

It is clear from table \ref{tab:table-2} that $\Xi$ is below 3\% for $0.9<\vartheta\leq 1.0$. Thus, in weak space charge limit, cosine law can be used in field emission module of a PIC simulation code. 

Another measure of space charge is the scaled current density $\zeta=kJ_PD^2V^{-3/2}$, where $J_P$ is the space charge affected current density due to field emission of electrons, $V=E_0 D$ is the diode-gap voltage and $k=\epsilon_0^{-1}\sqrt{m/2e}$ is a constant. For planar diode, $\vartheta$ and  $\zeta$ are related to each other as follows\cite{barbour,Forbes08,kyrit}:

\be
3\vartheta^2(1-\vartheta)=\zeta(4-9\zeta).
\label{eq:theta_zeta}
\ee  

\noindent
From the above equation, it may be seen that when $\vartheta \rightarrow 0$, the normalized current $\zeta\rightarrow 4/9$ i.e. $J_P\rightarrow J_{CL}$, where $J_{CL}=(4\epsilon_0/9)(\sqrt{2e/m}) (V^{3/2}/D^2)$ is the Child-Langmuir\cite{child,langmuir} current density.

In the weak space charge regime, $\vartheta \rightarrow 1$ and $\zeta \rightarrow 0$, the solution  to above equation around $\zeta=0$ may be expressed as:

\be
\vartheta=1-{\frac{4}{3}} \zeta + \ldots
\label{eq:purturbative}
\ee
\noindent
Note that in case of planar diode $\vartheta$ is ratio of electric field at the cathode in the presence of space charge and $E_0=V_g/D$. 

\begin{figure}[hbt]
  \begin{center}
 %\vskip -2.0cm
%\hskip  3.5cm
\hspace*{-0.7cm}
\includegraphics[scale=0.35,angle=0]{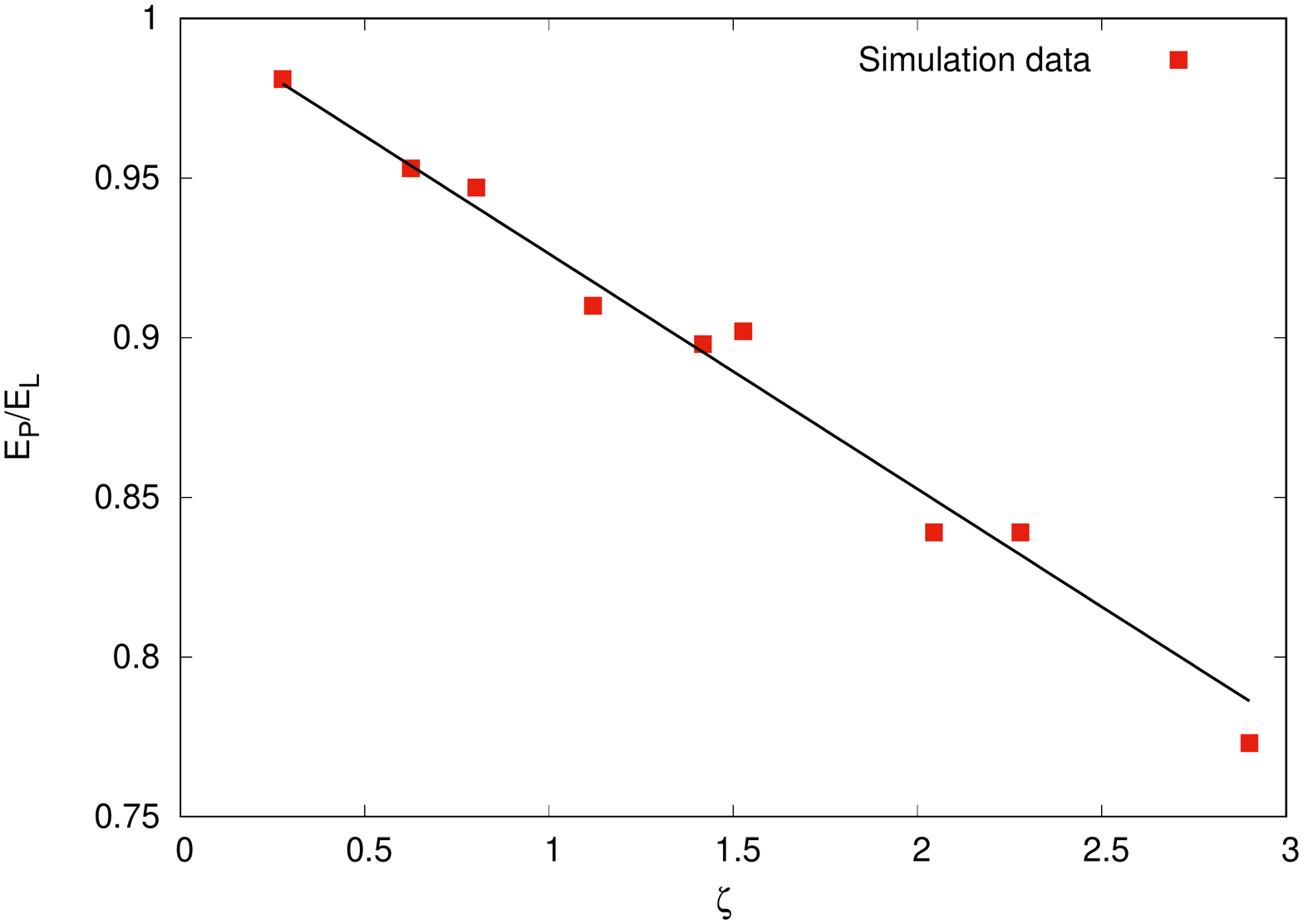}
\vskip -0.5cm
\caption{Normalized apex field $\vartheta=E_P/E_L$ plotted against normalized current $\zeta$ for $g=0.7$. Line $\vartheta=1.0-\delta \zeta$ was fitted to the points. Best fit gives  $\delta=0.073$. Best fitted line is also shown with the simulation point.}
\label{fig:theta_vs_zeta}
\end{center}
\end{figure}

In case of curved emitters, the relation between  $\vartheta$ and $\zeta$ may involve the field enhancement factor $\gamma_a$. This expectation stems from a recent result where it was shown that for curved emitters, the space charge limited current density scales as $J_{SCL} \simeq \gamma_a J_{CL}$\cite{PoP_PASUPAT}, here $\gamma_a =E_a/E_0$ is the field enhancement factor at the apex of the tip. Moreover, in case of emission from curved tips, the current density is not uniform over the emission area. Notwithstanding all these complications, we attempt to find a relation like $\vartheta=1-\delta\zeta$ by fitting the simulation data, where $\delta$ is a constant. However, calculation of $\zeta$ requires the current density $J_P$. To calculate $J_P$ from the current $I_P$ obtained in simulation, we need to divide $I_P$ by emission area $\cal A$. First we assume  ${\cal A}=\pi b^2$, i.e. base of the hemi-ellipsoidal tip. For this case, value of $\delta$ is found to be 0.43. But, field emission from curved surfaces is predominantly from the area in the vicinity of the apex of the tip. So a better way to calculate emission area may be to assume area  ${\cal A}=\pi (gR_a)^2$, where $g \leq 1$. Fig. \ref{fig:theta_vs_zeta} shows variation of $\vartheta$ with $\zeta$ obtained using PIC simulation. Here  $g$ was taken to be 0.7. A straight line $\vartheta=1 - \delta \zeta$  is fitted to the simulation points. We obtain $\delta=0.073$ for the best fit and the points appear to follow this linear trend even for $\vartheta < 0.9$. 

\begin{table}[h]
  \begin{center}
    \caption{Variation of fitting parameter $\delta$ with  $g$}
    \vskip 0.1 in
    \label{tab:table-3}
    \begin{tabular}{|c|c|c|c|c|c|c|}
      \hline
     $~g~$  &~0.5~ &~0.6~ &~0.7~ &~0.8~ &~0.9~ &~1.0~ \\
      \hline
      $~\delta~$ &0.037  &0.050 &0.073 &0.092 &0.12 &0.152          \\
       \hline
    \end{tabular}
\end{center}
\end{table}

Table \ref{tab:table-3} gives values of $\delta$ obtained for different values of $g$ corresponding to different emitting area at the tip. Note that values of $\delta$ obtained using simulation for all values of $g$ is quite small compared to $\delta=4/3$ predicted by  Eq.~\ref{eq:theta_zeta} which does not consider correction for the curved emitter. This needs further theoretical investigation with role of field enhancement factor considered properly.

\section{Conclusions}
\label{sec:conclusions}
The effect of space charge on the validity of the generalized cosine law of field variation at the surface of hemi-ellipsoidal emitter was investigated using PIC simulations. A special field emission module in PIC code PASUPAT was used for emission  of electrons. This module uses the cosine law of field variation to estimate the net current from the emitter. A distribution function based on cosine law is used to calculate the total energy, normal energy and launch angle of computational particles in the present study. It has been found that the average deviation $\Xi$ of the field from cosine law is less than 3\% for space charge parameter $\vartheta$ lying between 0.9 to 1.0. Thus for cases where space charge is not very strong, the field emission module based on cosine law can be used in PIC codes. As this scheme does not require calculation of current from each cell at the emitting surface, it helps in making the computation faster.   

We have also investigated the relation between normalized field at the apex of emitter, $\vartheta$ and normalized diode current $\zeta$ for curved emitter. In the weak space charge regime, simulation result shows that $\vartheta$ and $\zeta$ follow a linear relation similar to that for the flat emitters ($\vartheta=1 - \delta \zeta$). The value of $\delta$ is however small compared to that for the flat emitters. As the space charge limited current for curved emitters depends on the apex field enhancement factor, the theory for space charge affected  field emission may need to be modified for curved emitters. 

\vskip 0.170cm
{\it Acknowledgments}: R. K. thanks A. K. Poswal, Atomic and Molecular Physics Division, BARC for useful discussions. PASUPAT  simulations  were performed on ANUPAM-AGANYA super-computing facility at Computer Division, BARC.
\vskip 0.20cm
{ \it Data Availability}:  The data that supports the findings of this study are available within the article.

%\vskip 0.15cm

%\section{References} 

%\begin{references}

\appendix*

\section{Corrections to the distribution of emitted electrons}
\label{sec:appen}

In Ref.~[\onlinecite{db_parabolic}], the distribution of field-emitted electrons on the surface of a locally
parabolic sharp emitter was derived using the generalized cosine law. The sharpness assumption resulted in the
approximations $R_a/h \rightarrow 0$ and $z/h \rightarrow 1$ where $h$ is the height of emitter, $R_a$ the
apex radius of curvature and $z \approx h - \frac{\rho^2}{2R_a}$ is a point on the locally parabolic
surface of the axially symmetric emitter. These approximations are good to use when $h/R_a > 50$ but must be
replaced by more accurate calculations when the emitter is not so sharp.

The basic equation for the field emission current from a patch [$\rho,\rho + d\rho$] on the surface of
an axially symmetric emitter is

\be
dI  = f(\rho) d\rho = J_{MG}(\rho) 2\pi \rho \sqrt{1 + (dz/d\rho)^2} d\rho  \label{eq:Jbasic}
\ee

\noi
where $J_{MG}(\rho)$ is the Murphy-Good current density at ($\rho,z$) where $z \approx h -  \frac{\rho^2}{2R_a}$.
Since $J_{MG}(\rho)$ depends on the local field $E_l(\rho) = E_a \cos\tth$, where

\be
\cos\tth = \frac{\ttz}{\sqrt{\ttz^2 + \ttr^2}}, \label{eq:cos}
\ee

\noi
it is profitable to study the distribution of electrons with respect to $\tth$ rather than $\rho$. The scaled
variables $\ttz = z/h$ and $\ttr = \rho/R_a$ can be used to rewrite Eq.~(\ref{eq:Jbasic}) as

\be
dI  = f(\tth) d\tth = 2\pi R_a^2 J_{MG}(\tth) \ttr \sqrt{1 + \ttr^2} \frac{d\ttr}{d\tth}~ d\tth .\label{eq:Jtth}
\ee

\noi
In the following, we shall express $f(\tth)$ purely in terms of $\tth$ without resorting to the
approximations mentioned above.

We shall first express $\sqrt{1 + \ttr^2}$ in terms of $\sqrt{\ttz^2 + \ttr^2}$ without assuming $\ttz \approx 1$.
This can be easily achieved by noting that

\be
\begin{split}
  1 + \ttr^2 & = \ttz^2 + \ttr^2 + (1 - \ttz^2) =  (\ttz^2 + \ttr^2)\frac{1 + \ttr^2}{\ttz^2 + \ttr^2} \\
 =  & \left(\ttz^2 + \ttr^2\right) \left(\sin^2\tth + \frac{\cos^2\tth}{\ttz^2}\right) 
\end{split}
\ee

\noi
Thus, $\sqrt{1 + \ttr^2} = C_0 \sqrt{\ttz^2 + \ttr^2}$ where

\be
C_0  = \left(\sin^2\tth + \frac{\cos^2\tth}{\ttz^2}\right)^{1/2}.  \label{eq:C0}
\ee

The next correction requires an evaluation of $d\ttr/d\tth$. To achieve this, first
note that for a locally parabolic tip, $\ttr^2 \approx (1 - \ttz)2h/R_a$. Thus using Eq.~(\ref{eq:cos}),

\be
\ttz^2 (\sec^2\tth - 1) + \frac{2h}{R_a} \ttz - \frac{2h}{R_a} = 0
\ee

\noi
which can be solved to express $\ttz$ in terms of $\tth$ as

\be
\ttz = \frac{h}{R_a}\cot^2\tth \left[ \sqrt{1 + \frac{2R_a}{h}\tan^2\tth} - 1 \right].  \label{eq:Z}
\ee

\noi
Finally, using $\sin\tth = \ttr/\sqrt{\ttr^2 + \ttz^2}$, we have

\be
\cos\tth~ d\tth = \frac{d\ttr}{\sqrt{\ttr^2 + \ttz^2}} \left[ 1 -  \frac{\ttr^2 + \ttr \ttz \frac{d\ttz}{d\ttr}}{\ttr^2 + \ttz^2} \right]
\ee
  
\noi
where $d\ttz/d\ttr \approx -\ttr R_a/h$ for a locally parabolic tip. On simplification and using Eq.~(\ref{eq:Z}),

\be
\cos\tth~ d\tth =  \frac{\cos^2\tth}{\sqrt{\ttr^2 + \ttz^2}} C_1~  d\ttr
\ee

\noi
where

\be
C_1 = \sqrt{1 + \frac{2R_a}{h} \tan^2 \tth}.  \label{eq:C1}
\ee

Thus,

\be
\frac{d\ttr}{d\tth} = \frac{\sqrt{\ttr^2 + \ttz^2}}{\cos\tth} \frac{1}{C_1}.
\ee

Putting everything together and using the expressions for $\sin\tth$ and $\cos\tth$, we finally have

\bea
f(\tth)&  = & 2\pi R_a^2 J_{MG}(\tth) \ttr~\frac{\ttr^2 + \ttz^2}{\cos\tth} \frac{C_0}{C_1} \\
& = & 2\pi R_a^2 J_{MG}(\tth) \frac{\sin\tth}{\cos^4\tth}~\times \ttz^3 \frac{C_0}{C_1} \label{eq:final}
\eea

\noi
where the correction factor $\ttz^3 C_0/C_1$ can be evaluated using Eqns.~(\ref{eq:C0}), (\ref{eq:Z}),  and (\ref{eq:C1}).


\begin{thebibliography}{99}
\bibitem{tth} The angle $\tth$ closely approximates the angle $\theta_L$ that the local normal direction makes with respect to the emitter axis \cite{db_parabolic}.
\end{thebibliography}
\end{document}